\documentclass[prd,nofootinbib,twocolumn,showpacs]{revtex4}
\usepackage{graphics}
\usepackage{bm}

\def\be{\begin{equation}}
\def\ee{\end{equation}}
\def\ba{\begin{eqnarray}}
\def\ea{\end{eqnarray}}

\begin{document}

\title{Evolving dark energy equation of state and CMB/LSS cross-correlation}

\author{Levon Pogosian}

\smallskip

\affiliation{Institute of Cosmology, Department of Physics and Astronomy,
Tufts University, Medford, MA 02155 USA.}

\begin{abstract}
CMB power spectra and the SNIa luminosity-redshift curves have
difficulty distinguishing between models with the same average 
value of the dark energy equation of state. 
We propose using the cross-correlation of 
the CMB temperature anisotropy with large scale structure to help break this 
degeneracy.

\end{abstract}

\date{\today}

\maketitle

The evidence for Dark Energy (DE), a 
dark component causing the accelerated expansion of the universe,
comes from several complementary sources.
The analysis of the cosmic microwave background (CMB) shows that
the total energy density of the universe is very close to the critical
density, i.e. the universe it flat. At the same time, 
large scale structure (LSS) measurements suggest that no more 
than a third of the critical
energy density can be in the form of clustering matter.
While CMB and LSS measurements,  if considered separately,
can be explained without invoking DE,
they require DE for consistency with each other.
\cite{wmap_spergel,tegmark_etal}.
A more direct evidence for DE comes 
from measurements of the luminosity distance vs redshift for
supernovae type Ia (SNIa) \cite{Riess,Perlmutter} revealing an
accelerated expansion of the universe. 
For a flat Freedman-Robertson-Walker (FRW) universe acceleration implies 
domination by a substance with negative pressure, such as vacuum energy.
Another direct evidence which has recently become available is
the detection of the Integrated Sachs-Wolfe (ISW) effect using the
CMB/LSS cross-correlation \cite{xray,nolta,fosalba,scranton,afshordi03,vielva04,padma04}. 
This evidence is of complementary nature to the 
SNIa data. Rather than probing the accelerated expansion
it reflects the slow down in the growth of density perturbations expected 
in a non-matter dominated universe.

The cosmological constant $\Lambda$ is the
simplest DE model giving a satisfactory fit to the existing data.
Observations are also consistent with an evolving dark energy,
such as the scalar field Quintessence
\cite{wetterich,peebles,caldwell98}. Establishing whether the dark
energy is constant or evolving is one of the main challenges for
modern cosmology. In the FRW universe the large scale evolution of DE is determined
by its equation of state defined as the ratio of pressure to
energy density: $w \equiv p/\rho$. For scalar field Quintessence
$w$ also determines the clustering properties of DE. Depending on the
model $w$ can be constant or change with time. Models with $w \ne
-1$ correspond to evolving DE, while $w = -1$ corresponds to
$\Lambda$.

Much work has been done on trying to constrain the change in $w$
by fitting various forms of $w(z)$ to the SNIa luminosity distance-redshift data, 
often in combination with constraints from the CMB power spectrum. 
It is known, however, that these observables depend on $w(z)$ via one or 
more integrals over the redshift
\cite{huey99,maor01,maor_prd,dave02,ng,saini03}. As we illustrate below, the CMB power spectra
and the luminosity distance curves for models with varying $w(z)$
can be difficult to distinguish from predictions of the model with a constant
$w$ equal to the average of $w(z)$ defined as 
\be \label{averdef}
<w> \equiv {\int_{a_{*}}^1 da \ w(a) \ \Omega_D(a) \over
\int_{a_{*}}^1 da \ \Omega_D(a)} \ ,
\ee
where $a$ is the scale factor, $a=1$ today, $a_*$ is the scale factor at the
last scattering and $\Omega_D(a)$ is the ratio of the DE energy density
to the critical density.

The LSS measurements probe the net growth of cosmic structures.
Provided other parameters are known, LSS can differentiate
between a constant and a time-varying $w$, especially if one
had the information about the growth factor at 
different redshifts \cite{linder03}.

\begin{figure}[tbp]
\centering
\scalebox{0.4}{\includegraphics{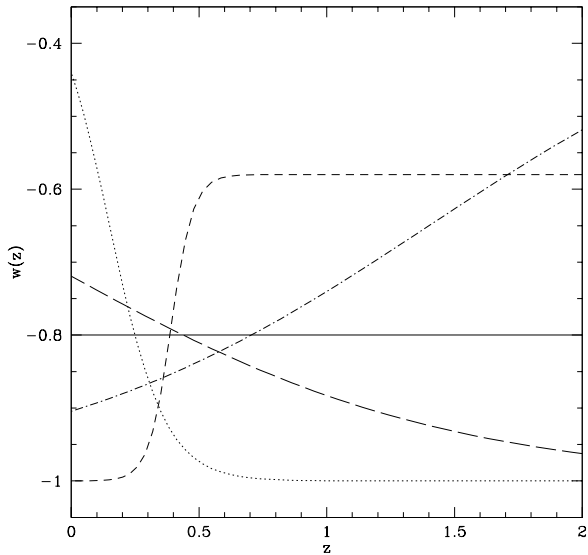}}
\caption{$w$ vs $z$ for the four varying $w(z)$ models considered in this paper. 
They all have the same average value $<w>$ (solid line) given by eq.~\ref{averdef}.}
\label{fig:eos}
\end{figure}

Measurements of the ISW effect can provide another probe of the
evolution of $w(z)$. In a way, they are more sensitive to the
change in $w(z)$ because ISW depends on the growth rate of
cosmic structures (the time-derivative of the growth factor) in addition to
the total growth (see \cite{CHB03} for a useful discussion). 
In this letter we build on the methods and results of \cite{GPV04} and 
illustrate the potential of the CMB/LSS cross-correlation for 
constraining the change in $w(z)$. For this purpose we will consider four 
Quintessence models with the same average $<w>$ shown in Fig.~\ref{fig:eos}. 
These four models are deliberately chosen
to illustrate a point and do not necessarily represent particularly
well-motivated quintessence potentials. The ansatz for $w(z)$ used to construct
our models is:
\be
w(z)= \left[{w_++w_- \over 2}\right] + \left[{w_+-w_- \over 2} \right] 
\ {\rm tanh}([z-z_0]/\delta z) \ .
\ee
It is similar in spirit to that of \cite{copeland03} and describes a transition 
from $w_+ \equiv w(z=+\infty)$ to $w_- \equiv w(z=-\infty)$
with parameters $\delta z$ and $z_0$ describing the width and the central redshift
of the transition. The values of the ansatz parameters used for the four
models in Fig.~\ref{fig:eos} are given in Table~\ref{table}.
The shapes of the corresponding Quintessence potentials
can be reconstructed from $w(z)$, if it was desired. These four models
have one thing in common -- they all have the same $<w>=-0.8$, as defined in
eq.~(\ref{averdef}). This value, $<w>=-0.8$, corresponds to the upper boundary of 
the $95\%$ confidence region obtained from the combined analysis of the WMAP, 
SDSS and SNIa data (Fig.~13 of \cite{tegmark_etal}).

\begin{table}[htbp]
\vskip 0.5 truecm
\begin{tabular}{|c||c|c|c|c|} \hline
{\bf Model / line type} \  & \ {\bf $w_-$ } \ & \ {\bf $w_+$ } \ & \ {\bf $z_0$ } \ & \ {\bf $\delta z$ } \ \\ \hline\hline 
$w=-0.8$, solid \ & \ -0.8 \ &  \ -0.8 \ & \ * \ &  \ * \ \\
dot \ & \ -0.3 \ &  \ -1 \ & \ 0.15 \ &  \ 0.22 \ \\
short dash \ & \ -1 \ &  \ -0.58 \ & \ 0.39 \ &  \ 0.09 \ \\
long dash \ & \ -0.4 \ &  \ -1 \ & \ -0.1 \ &  \ 1.55 \ \\ 
dot-dash \ & \ -1 \ &  \ -0.5 \ & \ 1. \ &  \ 1.98 \ \\ \hline
\end{tabular}
\caption{Models considered in the paper.}
\label{table}
\end{table}

\begin{figure}[tbp]
\centering
\scalebox{0.4}{\includegraphics{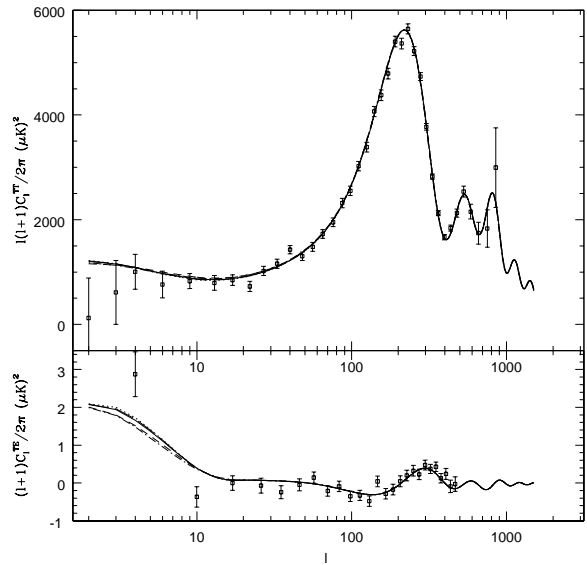}}
\caption{CMB temperature angular power spectrum (TT) and the temperature-polarization
cross-correlation (TE) for the five models in Fig.~\ref{fig:eos} (using the
same conventions for the different line types) with the WMAP's
first year data.}
\label{fig:cl}
\end{figure}

In Fig.~\ref{fig:cl} we show the CMB spectra for the models in Table~\ref{table}.
The cosmological 
parameters are the same for all models.
Through the paper we assume a flat
universe with the Hubble parameter $h=0.66$, baryon density $\Omega_b h^2=0.024$,
total matter density $\Omega_M h^2=0.14$, spectral index $n_s=0.99$, optical
depth $\tau=0.166$ and the amplitude of scalar fluctuations $A_s=0.86$ (as defined in
\cite{wmap_verde}.)
As evident from Fig.~\ref{fig:cl}, all four varying $w$ models fit the WMAP data 
very well and are indistinguishable from each other and from the model with a 
constant $w=-0.8$. This agrees with other studies (e.~g.~ \cite{corasaniti03})
and supports the point made in (among other papers) \cite{dave02} that
CMB spectra are mainly sensitive to the averaged value of $w$, as defined in eq.~(\ref{averdef})
\footnote{The choice of the parameterization of $w(z)$ has a major influence on the
strength of the constraints. E.~g., models with $w(z)$ chosen to be linear in $a$ or $z$
disfavor rapid variation \cite{jassal}. However, given sufficient freedom in a model, 
current data is perfectly consistent with a rapidly varying $w(z)$ \cite{corasaniti04,han04}.}.

\begin{figure}[tbp]
\centering
\scalebox{0.4}{\includegraphics{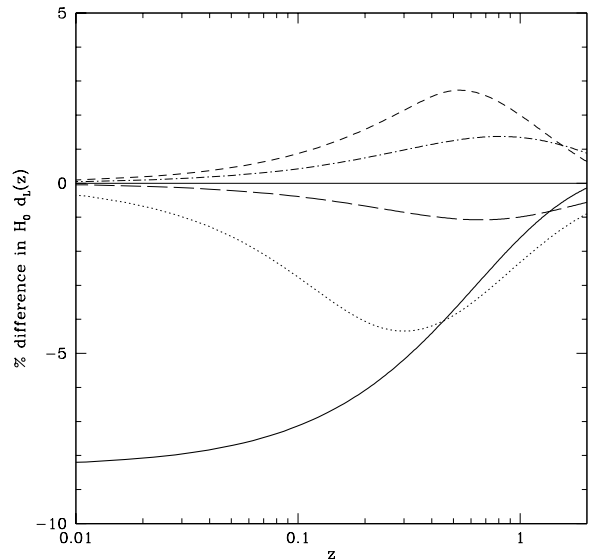}}
\caption{The percent difference in $d_L(z)$ for the
five models in Fig.~\ref{fig:eos} and the $\Lambda$CDM model (thick solid line).}
\label{fig:lumin}
\end{figure}

The degeneracy between the models persists to a large extent in the 
luminosity-redshift curves. The SNIa measurements probe the luminosity distance 
$ H_0 d_L(z)$, where
\be
d_L(z)=(1+z) \int_0^z {dz' \over H(z')} \ ,
\ee
$H(z)$ is the Hubble parameter and $H_0$ is its current value.
In Fig.~\ref{fig:lumin} we plot the percent difference between each of the
models with varying $w$ and the model of constant $w=-0.8$.
As can be seen from the figure, the differences from the $w=const$ case do not exceed $5\%$, 
which is just about, if not slightly below, the current level of 
accuracy with which $d_L(z)$ is extracted from SNIa. Therefore, a significantly varying 
$w(z)$ can be consistent with SNIa if the model has a reasonable average $<w>$. 
Most optimistically, future SNIa data can improve the accuracy
in determination of $d_L(z)$ to the level of $1-2\%$ \cite{maor_prd}. 
As the solid, dash-dot and long dash lines in Fig.~\ref{fig:lumin} show, 
the future SNIa data may be able to distinguish between an increasing and 
a decreasing $w(z)$, but there is still likely to be
a constant $w$ that gives a very similar $d_L$ for the same cosmological parameters.
This point was previously emphasized in \cite{maor01,maor_prd}.
Fig.~\ref{fig:lumin} also shows a clear difference in predictions of
the $w=-0.8$ model and the $\Lambda$CDM model. The $\Lambda$CDM model requires 
$h=0.72$ for the same $\Omega_M h^2$ in order to fit the CMB power spectra, hence
the $\Lambda$CDM curve in Fig.~\ref{fig:lumin} corresponds to a smaller $\Omega_M$.

\begin{figure}[tbp]
\centering
\scalebox{0.4}{\includegraphics{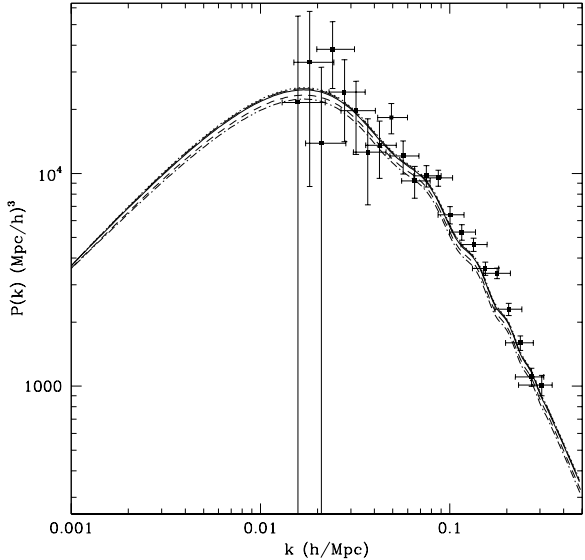}}
\caption{The linear matter power spectrum for the
five models in Fig.~\ref{fig:pk}. The SDSS data points \cite{sdss_pk} are plotted to show current 
experimental error bars.}
\label{fig:pk}
\end{figure}

Fig.~\ref{fig:pk} shows the matter power spectra for the models considered in the
paper. Plotted are
the linear CDM spectra at $z=0$. A proper accounting for the non-linear effects as 
well as for the redshift space distortion introduces non-negligible modifications
to the amplitudes and the shapes of the spectra. We did not attempt to do that here,
assuming that changes in the differences between the curves would be small. 
The SDSS data points are only plotted to show the uncertainty in the current 
determination of P(k). Not surprisingly, the models in which $w$ decreases with 
time are easier to distinguish, because the dark energy component becomes significant and
starts to affect the growth of fluctuations early. On the other hand,
in models with increasing $w(z)$ the growth of fluctuations is not
significantly affected until very recently and the change in $w$ with
time is harder to detect.
Knowing the matter power spectrum
at different redshifts
could provide a useful handle on the time-variation of $w$ \cite{linder03}.

If one assumes a flat universe with adiabatic initial conditions then the combination 
of the CMB with SNIa data provides the information about the cosmological parameters 
and $<w>$. The LSS data tightens the range of parameter values, fixes the galaxy bias 
and can give some 
information about the time variation of $w$.
Measurements of the CMB/LSS cross-correlation can play a complementary role and provide
further constraints on the time variation of $w$. What makes the cross-correlation
interesting for this purpose is that it depends not only on the total growth of 
cosmic structure but on the growth rate as well.

Cross-correlating the CMB with LSS with the purpose of detecting the ISW effect
was first proposed in \cite{turok96}.
Let us define
\begin{equation}
\Delta(\hat{\bf n}) \equiv T(\hat{\bf n})- \bar{T}
\end{equation}
and
\begin{equation}
\delta(\hat{\bf n}) \equiv 
{\rho(\hat{\bf n})- {\bar\rho} \over {\bar\rho}} \ ,
\end{equation}
where $T(\hat{\bf n})$ is the CMB temperature measured along the direction 
$\hat{\bf n}$,
$\rho(\hat{\bf n})$ is the mass density along $\hat{\bf n}$, and
$\bar{T}$ and $\bar{\rho}$ are the averaged CMB temperature and the
matter density.
The temperature anisotropy due to the ISW effect is an integral over the
conformal time:
\begin{equation}
{\Delta(\hat{\bf n}) \over \bar{T}}= \int_{\eta_r}^{\eta_0} d\eta 
( \dot{\Phi}-\dot{\Psi}) \left[(\eta_0-\eta) \hat{\bf n},\eta \right] ,
\end{equation}
where $\eta_r$ is some initial time deep in the radiation era, $\eta_0$
is the time today, $\Phi$ and $\Psi$ are the Newtonian gauge gravitational 
potentials,
and the dot denotes differentiation with respect to $\eta$.
The quantity $\delta(\hat{\bf n})$ contains contributions from astrophysical
objects (e.~g. galaxies) at different redshifts and can also be expressed as an 
integral over the conformal time:
\begin{equation}
\delta(\hat{\bf n}) = \int_{\eta_r}^{\eta_0} d\eta \ {dz \over d\eta} 
\ W_g(z(\eta)) \ 
{\delta}((\eta_0-\eta) \hat{\bf n},\eta) ,
\end{equation}
where $W_g(z)$ is a normalized galaxy selection function. We do not explicitly
include a possible bias factor, assuming it is determined from the measurements
of the LSS.
We are interested in calculating the cross-correlation function
\begin{equation}
X(\theta) \equiv X(|\hat{\bf n}_1-\hat{\bf n}_2|) \equiv 
\langle \Delta(\hat{\bf n}_1) \delta(\hat{\bf n}_2) \rangle \ ,
\end{equation}
where the angular brackets denote ensemble averaging and $\theta$ is
the angle between directions $\hat{\bf n}_1$ and $\hat{\bf n}_2$.
It can be written as \cite{GPV04}
\begin{eqnarray}
X(\theta) &=& {9 \over 25} \int_{\eta_r}^{\eta_0} d\eta_1 
\int_{\eta_r}^{\eta_0} d\eta_2 \ 
\dot z(\eta_2) \ W_g(z(\eta_2)) \nonumber \\ &\times&
\int {dk \over k} \Delta_{\cal R}^2(k) \ {{\rm sin}(kR)\over kR} 
\ F(k,\eta_1,\eta_2) \ ,
\nonumber \\ {}
\label{wtg3}
\end{eqnarray}
where $k$ is the wave-number, $\Delta_{\cal R}^2(k)$ is the primordial curvature 
power spectrum, $F(k,\eta_1,\eta_2)$ describes the time-evolution of 
$(\dot{\Phi}-\dot{\Psi})$ and $\delta$, 
$R\equiv \sqrt{r_1^2+r_2^2-2r_1r_2{\rm cos}\theta}$ and $r \equiv \eta_0 - \eta$ 
(see \cite{GPV04} for more details) 
\footnote{In addition, one has to subtract the monopole and dipole contributions 
to eq~(\ref{wtg3}). The details of how it was done can be found in \cite{GPV04}.}. 

\begin{figure}[tbp]
\centering
\scalebox{0.4}{\includegraphics{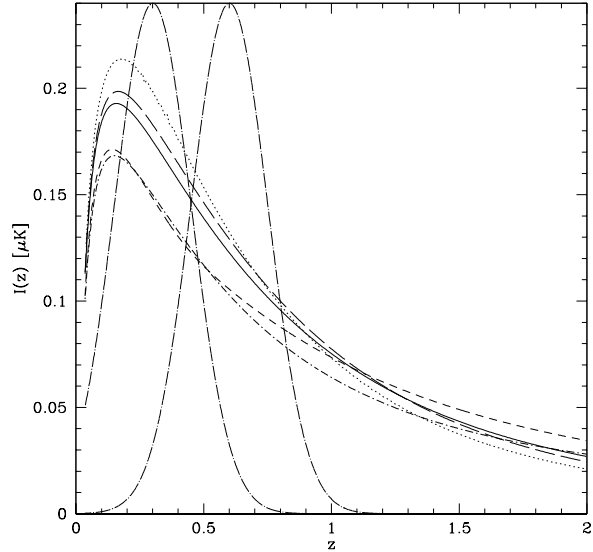}}
\caption{The integrand $I(z,\theta=0.2^\circ)$ for the
five models in Fig.~\ref{fig:pk}. Plotted on top  (long dash-dot) are the two 
choices of window functions used in the paper, rescaled to fit the box.}
\label{fig:int}
\end{figure}

The galaxy window function $W_g(z)$ plays an important role as it essentially
zooms in on the ISW effect over a particular redshift range. To help one
decide on the selection of $W_g(z)$ it is useful to define 
$I(\theta,z)$ as
\be
I(\theta,z) \equiv {9 \over 25} \int_{\eta_r}^{\eta_0} d\eta_1 
\int {dk \over k} \Delta_{\cal R}^2(k) \ {{\rm sin}(kR)\over kR} 
\ F(k,\eta_1,\eta_2) \ , 
\label{integrand}
\ee
so that eq.~(\ref{wtg3}) can be written as
\begin{eqnarray}
X(\theta) &=&  \int_{z_r}^{0} dz \ W_g(z) I(z,\theta) \ .
\end{eqnarray}
Typically, the theoretical prediction for the
cross-correlation approaches a flat plateau for $\theta < 0.2^\circ$ and
the hight of the plateau is a good measure of the overall signal.
In Fig.~\ref{fig:int} we plot $I(z,\theta=0.2^\circ)$ vs $z$ for the
models considered in the paper. From this plot one can guess
where to place the window function to see
the maximum difference in $X(\theta)$ for the models. This, however, may
not correspond to the choice that gives the highest signal-to-noise ratio.
Due to the arbitrary nature of the models considered in the paper
we did not search for the optimal window function(s) that 
would maximize the difference between the models while minimizing the statistical
uncertainty in $X(\theta)$. Such optimization procedure, however, would be
necessary if the cross-correlation was used to study properties of dark energy.

\begin{figure}[tbp]
\centering
\scalebox{0.4}{\includegraphics{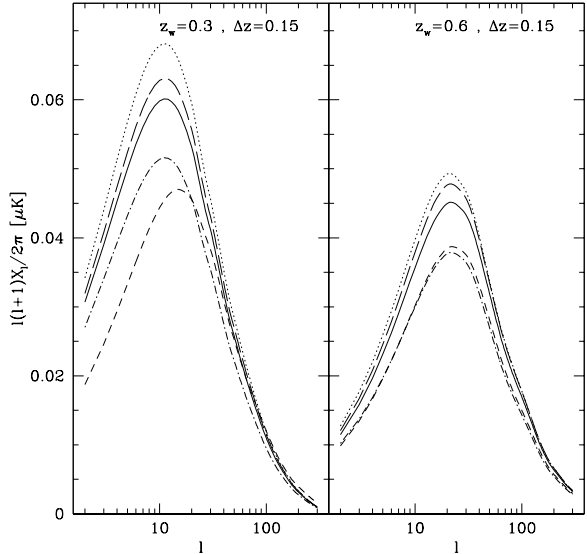}}
\caption{The angular cross-correlation, $l(l+1)X_l/2\pi$ for the
five models in Fig.~\ref{fig:pk} and two choices of window functions.
Theoretical uncertainties in individual $X_\ell$, if plotted, would be well outside 
the boundaries of the box. Instead, we propose using
the cumulative cross-correlation (summed over $\ell$) to differentiate between 
the models.}
\label{fig:xl}
\end{figure}

For computational purposes it is advantageous to decompose $X(\theta)$ 
into a Legendre series,
\begin{eqnarray}
X(\theta)= 
\sum_{l=2}^{\infty} {2\ell+1 \over 4\pi} X_\ell P_\ell({\rm cos} \ \theta) \ .
\end{eqnarray}
In Fig.~\ref{fig:xl} we show results
for the angular cross-correlation $l(l+1)X_\ell/2\pi$ for the two choices of Gaussian
window functions shown in Fig.~\ref{fig:int}. One is centered at $z_w=0.3$ and 
the other at $z_w=0.6$, both have a standard deviations of $\Delta z =0.15$.
The two window functions are chosen so that the first one results in the larger difference between
the models, while the second one gives the larger signal-to-noise ratio.
Theoretical uncertainties in individual $X_\ell$ are too
large to allow comparing the models at each $\ell$.
Instead one could calculate the total signal, $X_T$, for each model by
summing $(2l+1)X_\ell/4\pi$ over $\ell$:
\be
X_T = \sum_{\ell=2}^{\ell_{\rm max}} {2\ell+1 \over 4\pi}X_\ell \ .
\ee 
Note that, for $l_{\rm max} = \infty$, $X_T=X(\theta=0)$.
The best achievable signal-to-noise in determination of $X_T$ is approximately given by
\be
\left({S \over N}\right)^2 \approx f_{\rm sky} \sum_{\ell=2}^{\ell_{\rm max}}(2\ell+1)
{X_\ell^2 \over C_\ell^{TT}C_\ell^{MM}} \ ,
\ee
where $f_{\rm sky}$ is the surveyed fraction of the sky, $C_\ell^{TT}$ is the
angular CMB temperature power spectrum and $C_\ell^{MM}$ is the analogously
defined matter angular spectrum inside a given window $W_g(z)$.
Table \ref{xt} contains $X_T$ for each model for the two choices of the
window functions.
\begin{table}[htbp]
\vskip 0.5 truecm
\begin{tabular}{|c||c|c|} \hline
{\bf Model/line type} \  & \ {\bf $X_T(z_w=0.3)$} \ & \ {\bf $X_T(z_w=0.6)$} \ \\ \hline\hline 
$w=-0.8$, thick solid \ & \ 0.167 \ &  \ 0.125 \  \\
dot \ & \ 0.187 \ &  \ 0.137 \ \\
short dash \ & \ 0.132 \ &  \ 0.108 \ \\
long dash \ & \ 0.175 \ &  \ 0.132 \ \\ 
dot-dash \ & \ 0.144 \ &  \ 0.106 \ \\ \hline
\end{tabular}
\caption{The cumulative cross-correlation $X_T$, in $\mu K$, for $\ell_{\rm max}=300$ for 
the models considered in the paper and the two choices of window functions.}
\label{xt}
\end{table}
The table shows differences of up to $\sim 35\%$ in $X_T$ for the window function 
centered at $z_w=0.3$ and up to $\sim 25\%$ for the one centered at $z_w=0.6$.
The signal-to-noise does not significantly depend on the particular model. Typically, 
$S/N \approx 2$ for the window function with $z_w=0.3$ and $S/N \approx 3$ for $z_w=0.6$, 
assuming complete
sky coverage ($f_{\rm sky}=1$) and that $C_\ell^{TT}$ contains not just the ISW contribution
but the entire CMB anisotropy on relevant scales.
Neither choice of the window function would distinguish between the models 
considered in the paper. There is hope, however, that the uncertainty in
$X_T$ can be reduced to the level that would allow model comparison.

The potential accuracy that can be achieved in cross-correlation measurements was 
studied in \cite{cooray,afshordi04,scranton04}. The main theoretical uncertainty
arises from limitations imposed by cosmic variance and
the fact that, in addition to the ISW contribution, the CMB signal
has a sizable primary component.
A possible way to increase the
signal to noise, proposed in \cite{afshordi04}, is to consider the
cross-correlation in multiple redshift shells. If the shells could be
selected in a way that they were nearly uncorrelated then one could
treat the cross-correlation in each shell as a separate measurement.
Based on the results of \cite{afshordi04} it appears that the best 
accuracy one could hope for is about $10\%$. If this was the case, 
the cross-correlation 
could be useful for adding to the constraints on rapidly varying $w(z)$.
The CMB/LSS correlation could also help with constraining the clustering properties
of Dark Energy (see e.g. \cite{scranton04,ps05}).

There is still room for a better understanding of the inherent limitations of the
cross-correlation analysis. In particular, there is need for
a procedure that would optimize the selection of redshift shells to
maximize the signal to noise. This, as well as the constraints from the existing data 
on the variation of $w(z)$, is a subject of ongoing work \cite{RobBobP}.

\acknowledgments
I would like to thank Ramy Brustein, Rob Crittenden, Jaume Garriga, 
Dragan Huterer, Bob Nichol and Tanmay Vachaspati for very helpful discussions
and comments.
The code used for calculating the cross-correlation was developed in
collaboration with Jaume Garriga and Tanmay Vachaspati \cite{GPV04} and was based on
CMBFAST \cite{cmbfast}.

\end{document}